\documentstyle[11pt,newpasp,twoside,epsf]{article}
\def\edcomment#1{\iffalse\marginpar{\raggedright\sl#1\/}\else\relax\fi}
\reversemarginpar

\begin{document}
\title{Early results from the VIMOS VLT Deep Survey}
\author{O. Le Fevre$^1$, G. Vettolani$^2$, D. Maccagni$^3$, 
J.P. Picat$^4$,  C. Adami$^1$, M. Arnaboldi$^5$, S. Bardelli$^{10}$,
M. Bolzonella$^{10}$,
M. Bondi$^2$, D. Bottini$^3$, G. Busarello$^5$, A. Cappi$^2$, 
P. Ciliegi$^2$, T. Contini$^4$,
S. Charlot$^7$, S. Foucaud$^3$,  P. Franzetti$^3$,
B. Garilli$^3$, I. Gavignaud$^4$, L. Guzzo$^8$, O. Ilbert$^1$, 
A. Iovino$^8$, V. Le Brun$^1$, B. Marano$^2$, C. Marinoni$^1$,
H.J. McCracken$^2$, G. Mathez$^4$, A. Mazure$^1$,
Y. Mellier$^6$, B. Meneux$^1$
P. Merluzzi$^5$, S. Paltani$^1$,
R. Pell\`o$^4$, A. Pollo$^8$, L. Pozzetti$^2$,
M. Radovich$^5$, D. Rizzo$^8$, R. Scaramella$^9$, 
M. Scodeggio$^3$, L. Tresse$^1$,
G. Zamorani$^{10}$, A. Zanichelli$^2$, E. Zucca$^{10}$}
\affil{$^1$Laboratoire d'Astrophysique de Marseille - France, 
$^2$Istituto di Radio-Astronomia - CNR, Bologna - Italy,
$^3$IASF - INAF, Milano - Italy, $^4$Laboratoire d'Astrophysique - Observatoire
Midi-Pyr\'en\'ees,
$^5$Osservatorio Astronomico di Capodimonte, Naples, Italy
$^6$Institut d'Astrophysique de Paris, Paris, France,
$^7$Max Planck fur Astrophysik, Garching, Germany
$^8$Osservatorio Astronomico di Brera, Milan, Italy,
$^9$Osservatorio Astronomico di Roma, Italy,
$^{10}$Osservatorio Astronomico di Bologna, Bologna, Italy}

\begin{abstract}
The VIMOS VLT Deep Survey (VVDS) is underway to study the
evolution of galaxies, large scale structures and AGNs, from
the measurement of more than 100000 spectra of faint
objects. We present here the results from the 
first epoch observations of more than 20000 spectra. 
The main challenge of the program, the redshift measurements,
is described, in particular entering the ``redshift desert''
in the range $1.5<z<3$ for which only 
very weak features are observed in the observed wavelength range.
The redshift distribution of a magnitude limited sample
brighter than $I_{AB}=24$ is presented for the first time,
showing a peak at a low redshift $z\sim0.7$, and a tail 
extending all the way above $z=4$. The evolution of the 
luminosity function out to $z=1.5$ is presented, with the
LF of blue star forming galaxies carrying most of the 
evolution, with L* changing by more than two magnitudes
for this sub-sample.
\end{abstract}

\section{Introduction}

Understanding how galaxies and large scale structures
formed and evolved is one of the major goals of modern cosmology.
In order to identify the relative contributions of the various
physical processes at play and the
associated timescales, a comprehensive
picture of the evolutionary properties of the constituents of the universe
is needed over a large volume and a large time base. Samples of high redshift
galaxies known today reach less than a few thousand at redshifts $1-3$, and
statistical analysis suffer from small number statistics, small
explored volumes, selection biases, which prevent detailed analysis.
In the local universe, large surveys like the 2dFGRS and the SDSS,
will contain from $2.5\times10^5$ up to $10^6$ galaxies to reach
a high level of accuracy in measuring the fundamental parameters
of the galaxies and AGNs populations. Similarly, we need to
gather large numbers of galaxies at high redshifts to accurately
describe the various populations in environments
ranging from low density to the dense cluster cores, and relate
the properties observed at different redshifts to identify
the main processes driving evolution. 

This goal can only be achieved through massive observational
programs assembling galaxies and QSO samples representative of 
the universe at the various look-back times explored. As the 
observed galaxies are at large distances and therefore
very faint, instruments have to be conceived to combine
wide field, high throughput, and high multiplex gain in order
to efficiently observe large samples. 
Multi-object spectrographs are routinely in operation
since $\sim$15 years, and the new generation now in
place on the $6-10$m telescopes, like DEIMOS on the Keck-10m
or VIMOS on the VLT-8m, allows to explore large
volumes of the distant universe through the observations
of many tens of thousand of objects. 

Because of this leap forward in instrument performances, several
large deep surveys are now in progress. We are presenting
here the VIMOS VLT Deep Survey (VVDS). It is combining
a common observing strategy to assemble more than 100000
redshifts in 3 large galaxy datasets observed over more than
16deg$^2$ in 5 equatorial fields, selected in magnitude
from $I_{AB}\leq22.5$ to $I_{AB}\leq25$. The general strategy
of the survey has been presented in e.g. Le F\`evre et al. (2001).

\section{Survey observations status}

The first survey observations have been conducted during the
dark time of November and December 2002. A total of 29 nights
have been allocated from the garanteed time, out of which 18 have
been clear. The survey priority was set to accumulate as much
observations as possible on the VVDS-0226-04 deep field,
to a depth $I_{AB}=24$. The remaining time was spend on the
2217+00 and the 1002+03 fields, as part of the VVDS-wide 
survey. Targets have been selected from the deep imaging
survey we have conducted in these fields (Le F\`evre et al.,
2003). The VIMOS observations have been performed using
1 arcseconds wide slits and the LRRed grism. This provides
a spectral resolution of $R=210$. The length of spectra
at this resolution allows to pack on average 350--400 spectra
for the ``wide'' $I_{AB}\leq22.5$ survey on the 4 VIMOS detectors, 
and 500 to 600 spectra (figure 1) for the ``faint'' $I_{AB}\leq24$, 
a key factor to accumulate
redshifts in a very efficient way.

For a magnitude limited sample potentialy going to
redshifts up to $\sim5$, the choice of the observed 
wavelength range is difficult: too blue and the 
rest frame domain around OII-4000$\AA$ break which contains
strong features, disappears
from the band already for $z\sim1$, too red and the 
UV domain comes into the band for quite high redshifts. 
We have compromised with a wavelength domain $5500-9400\AA$,
which allows to follow [OII] up to $z=1.5$, at the expense
of having to deal with the faint absorption features present in the
rest frame domain corresponding to the redshift 
range $1.5<z<3$.

A total of 49 pointings (one pointing is a set of 4 masks
in each of the 4 VIMOS quadrants) have been observed, and a 
total of 20741 spectra have been obtained, some objects
being observed in two different pointings to measure
the redshift measurement errors associated to the observational
process.

\begin{figure}  
%\plotone{1001_galaxies.ps}
\plotfiddle{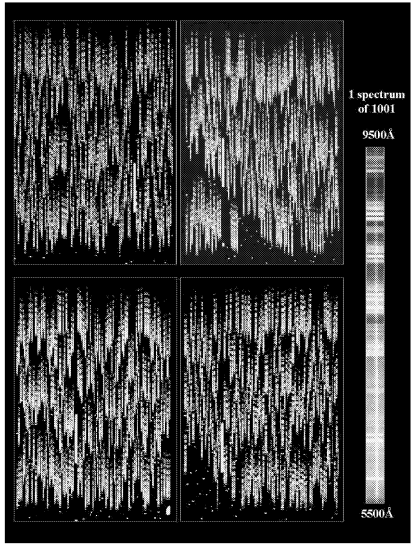}{7cm}{0}{31}{31}{-100}{-15}
\caption{VIMOS-VLT observations of 1001 galaxies, demonstrating the
high multiplex capability of the instrument}
\end{figure}

\section{Data processing and redshift measurement}

The data processing is conducted in two steps. First the large number
of raw data frames are organised in the object data and various
calibration categories. This is followed by 2D spectra extraction, 
sky correction, sum/combine, and then by 1D optimal spectra extraction,
wavelength and flux calibration. This is performed very
efficiently with the
VVDS consortium pipeline VIPGI with a minimum of human
interaction (Scodeggio et al.,
in preparation). 

The KBRED environment developed
under IDL is then used to compute redshifts (Scaramella 
et al., in preparation). KBRED is run automatically
on the whole sample, followed by systematic examination
of each single spectrum by eye in the VIPGI environment.
KBRED basically works by cross-correlation of observed 
spectra with a set of library template spectra 
followed by a reconstruction of each spectrum on
a base of template vectors (Principal Component Analysis, PCA).
There are two main limitations for this process
to be fully automated.
Because the range of possible redshift solutions to explore
is large, with objects as diverse as stars, galaxies with redshifts
possibly up to z=5 or more, or AGNs, it is essential to
assemble as complete as possible a set of star templates,
galaxies and AGN templates covering
the full rest frame wavelength range from Ly$\alpha-1216$
to H$\alpha-6562\AA$, for early to late spectral types. 
Because of the lack of 
observed template spectra in the literature 
in the range between CIV-1548 and OII-3727,
critical for redshifts $1.5-3$ given our wavelength
range, we had to devise a scheme to go through
this so called ``redshift desert'', which is obviously only
produced by the observational bias of spectroscopy set-ups. 
We have taken an iterative approach, creating templates
from VIMOS spectra, and updating them along the reduction
process as new galaxies are identified at redshifts
$z>1.3$. This process is helping to
significantly reduce the first-pass incompletness, and 
we are identifying several hundred galaxies beyond z=1.5
(Le F\`evre et al., in preparation).
The second limitation is linked to the residuals of 
the sky subtraction and fringing correction still present
in the reduced spectra which produce a noise
pattern highly non linear, variable from spectra to
spectra. Eye examination 
of both the 2D and 1D spectra are therefore sometimes required
to validate a redshift proposed by KBRED. We are currently
working at computing a two 2D noise model in each
slit to help reduce this limitation.

Example spectra processed through VIPGI and KBRED are shown
in Figures 6 and 7.

\section{Redshift distribution to $I_{AB}=22.5$ and $I_{AB}=24$}

The redshift distribution of the $I_{AB}=24$ sample 
is presented in Figure 2, the first time such a
distribution is observed to this depth. At the time of this 
writing, this distribution is affected by incompletness above 
$z\sim1.5$, waiting for a complete treatment of the
as yet unidentified spectra as described in the previous section. 
However, there are two main features to be noted: the mean
redshift is $<z>=0.7$, and there is a small
but significant high redshift tail of galaxies identified 
up to z=5. In comparison, our $I_{AB}=22.5$ sample
shows a mean redshift of $<z>=0.6$ (Figure 2).
Most semi-analytic models make predictions for
a higher mean redshift and a more numerous high redshift
tail above z=1.5 for a $I_{AB}=24$ limited sample. While 
our current incompletness is most probably caused by
objects at $z>1$ as identified from photometric redshifts
estimates, unidentified objects cannot significantly
affect the mean redshift of the full sample. This is 
dicussed in more details in Le F\`evre et al. (in preparation).

\begin{figure}  
\plottwo{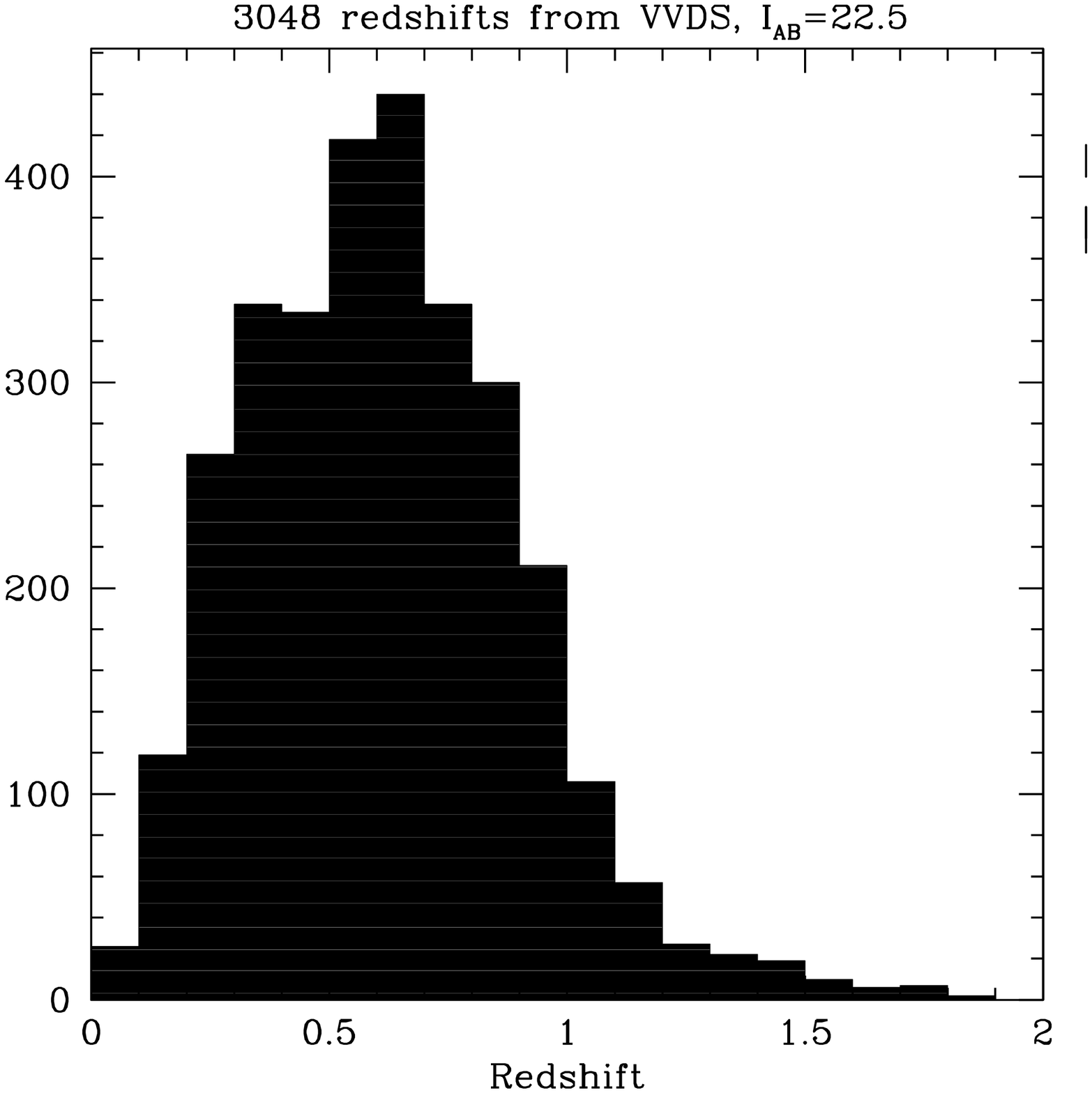}{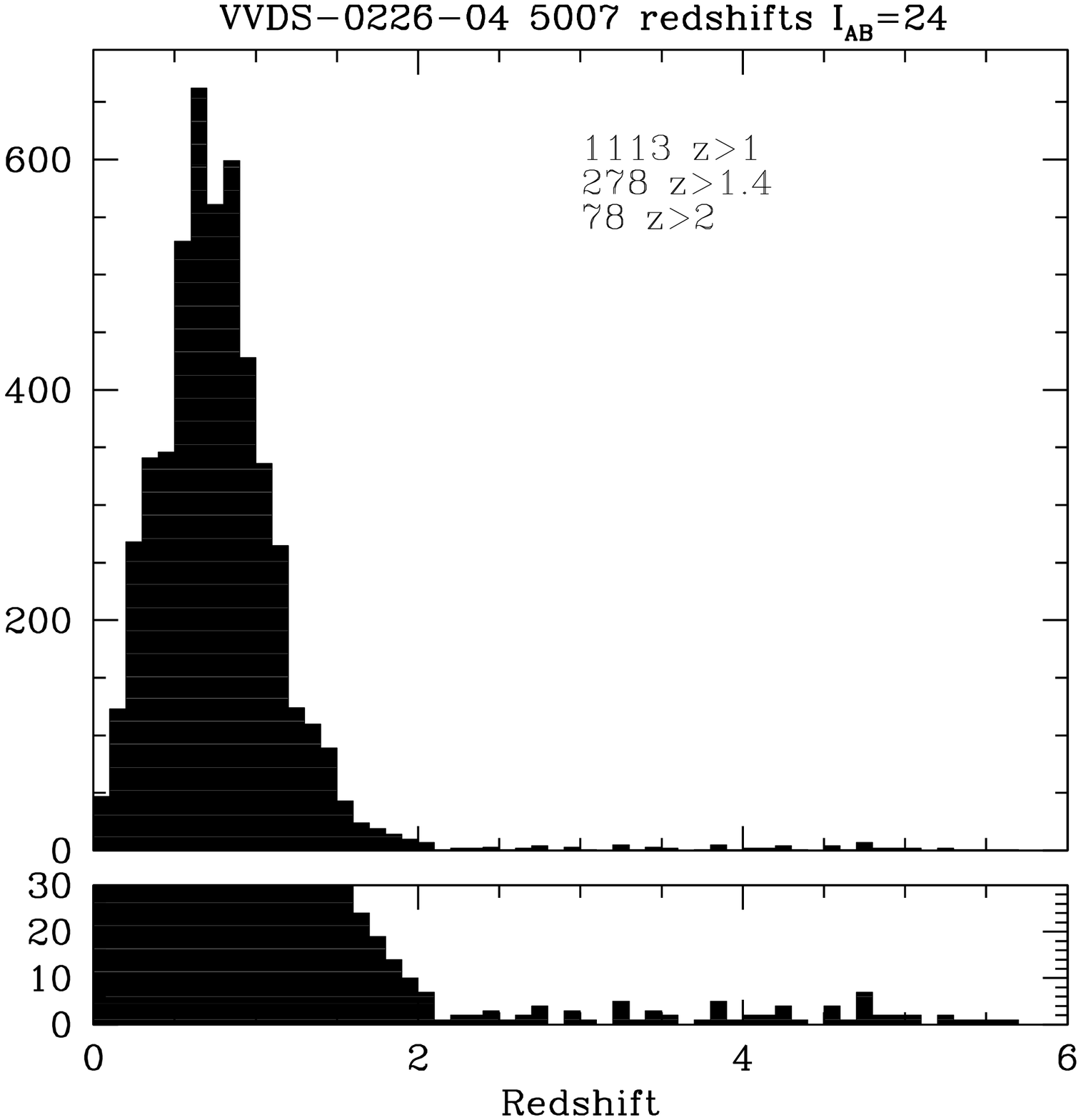}
%\plotfiddle{zdist_f02.ps}{7.5cm}{0}{40}{40}{-130}{-70}
\caption{{\it (Left)} Redshift distribution for a magnitude limited sample 
of 3038 galaxies with $I_{AB}\leq22.5$ from the VVDS ``wide'', and 
{\it (Right)} 5007 galaxies with $I_{AB}\leq24$ in the VVDS-0226-04 
``deep'' field
as measured in September 2003 (preliminary).}
\end{figure}

\section{Evolution of the luminosity function}

With the current sample, we can compute the 
luminosity function out to $z\sim1.5$ although our
current incompletness starts to kick in
at $z>1.2$. Figure 3 shows the 
luminosity function of a $I_{AB}=24$ sample of
4015 galaxies, as a function of redshift out to
z=1.5. There is an overall brightening of the luminosity
by about 1 magnitude at $z\sim1$, with a strong steepening of
the faint end slope clearly identified (see Zucca et al.,
in preparation). 

\begin{figure}  
%\plotone{figLF.ps}
\plotfiddle{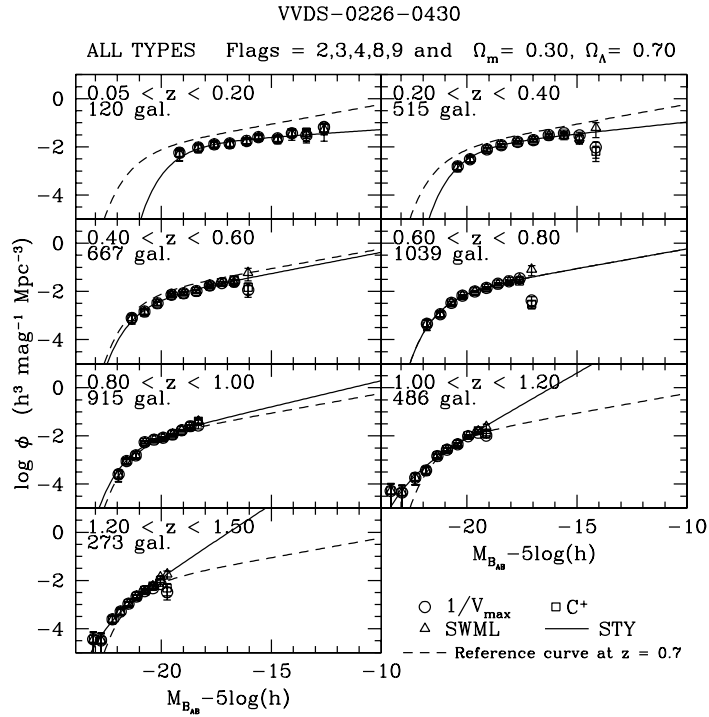}{8.5cm}{0}{47}{47}{-150}{-70}
\caption{Global luminosity function (preliminary) build from a sample of
4015 galaxies in the VVDS-0226-04 field, as a function of 
redshift. The dashed line in each panel 
represents the LF fit in the 0.6-0.7 redshift range.}
\end{figure}

Since we have sufficient number statistics, we can for
the first time break up the evolution of the LF by galaxy 
spectral types with the current data set reliably
up to redshifts $z\sim1.2$ and $M_{B_{AB}}\leq-19$.
Each galaxy can be assigned a spectral type based on the 
SED model fitting of the photometric data (covering a larger
wavelength base than the spectra themselves). With all galaxies
classified in 4 types, the LF build for the 
two extreme types of the sample, early types and blue star forming
spectral types is shown in Figure 4 \& 5.
While the LF of the early type population seems not to
be evolving my more than 0.5 mag., the LF of the blue star forming 
galaxies is strongly evolving with redshift, with the L* 
luminosity measured from the Schechter function fit increasing
by more than two magnitudes by redshift $\sim1$.
This strong differential evolution vs. spectral type
is discussed in Ilbert et al., in preparation. 

We will aim
to describe the evolution of the population vs. type 
as well as vs. local galaxy density when the processing
of the current sample is complete.

\section{Other VVDS products}

The luminosity density and the 
star formation rate history are being produced from the
luminosity functions. We will be computing the
correlation function $\xi(r_p,\pi,z)$ and $w(r_p,z)$ projections
for the full sample as well as per each galaxy spectral type
to map the evolution of the clustering vs. redshift.

Several other data products are directly available from the
VVDS because it is a magnitude limited survey. 
About 100 type 1 AGNs have been 
identified in the current sample, the highest redshift QSO being
at $z=5.000$ (Figure 5), this faint AGN sample is highly complementary to
other surveys to date since it is surveying AGNs $2-3$ magnitudes deeper.
Several clusters of galaxies are being identified
and their properties studied. 

The high redshift population observed in the VVDS is a unique
sample assembled from a purely magnitude selected photometric sample.
This galaxy census, when complete, will enable to derive evolution
out to $z\sim4$ {\it from within the same, large, homogeneous sample}.

The VVDS is extending to multiwavelength observations:
we have been conducting a
deep radio survey (Bondi et al., 2003), and are associated with
other teams conducting observations at other wavelengths:
Galex, XMM, SIRTF. Furthermore, the VVDS-10h field is now the target 
of the COSMOS program with a 640 orbits HST-ACS coverage
of 2 deg$^2$, with extensive multi-wavelength coverage (P.I.:
N. Scoville).

\begin{figure}  
%\plottwo{LF_Type1_B.ps}{LF_Type4_B.ps}
\plotfiddle{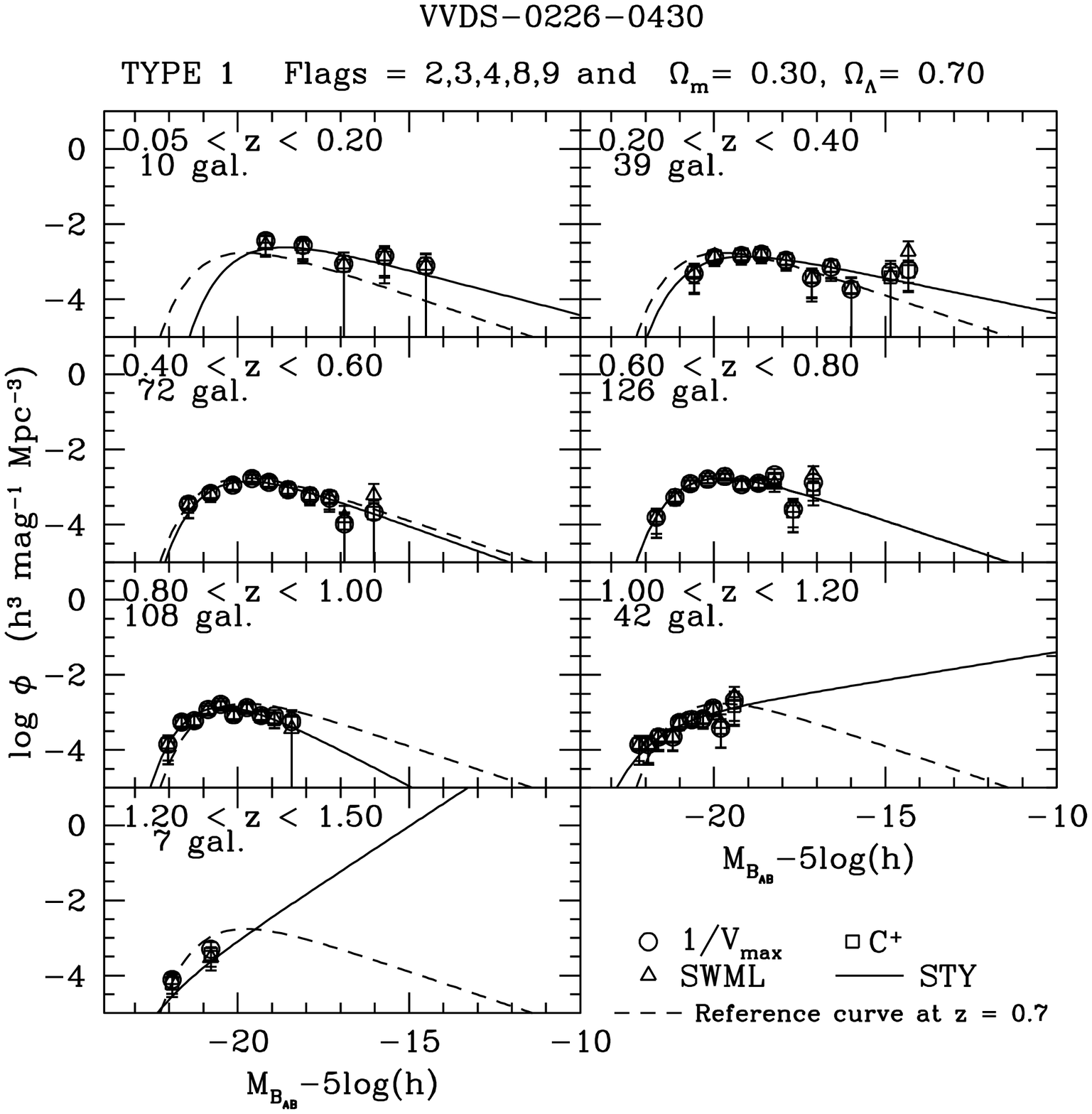}{7.5cm}{0}{45}{45}{-150}{-70}
\caption{Preliminary Luminosity function build from the sample of
404 early type {\it left)} and 1665 actively star forming galaxies 
{\it (right)} in the VVDS-0226-04 field.}
\end{figure}

\begin{figure}  
%\plottwo{LF_Type1_B.ps}{LF_Type4_B.ps}
\plotfiddle{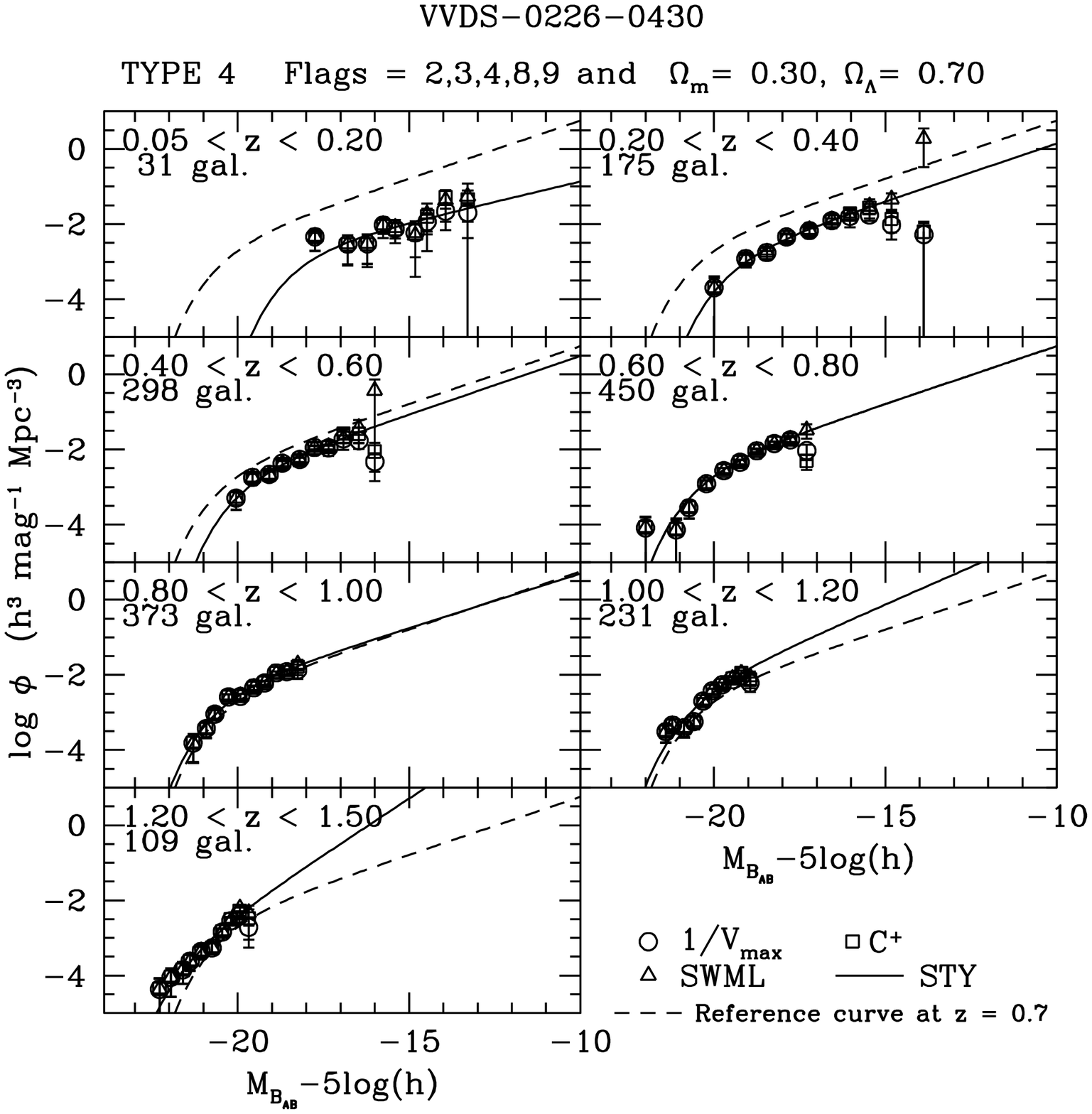}{7.5cm}{0}{45}{45}{-150}{-70}
\caption{Preliminary Luminosity function build from the sample of
404 early type {\it left)} and 1665 actively star forming galaxies 
{\it (right)} in the VVDS-0226-04 field.}
\end{figure}

\begin{figure}  
%\plotone{test.ps}
\plotfiddle{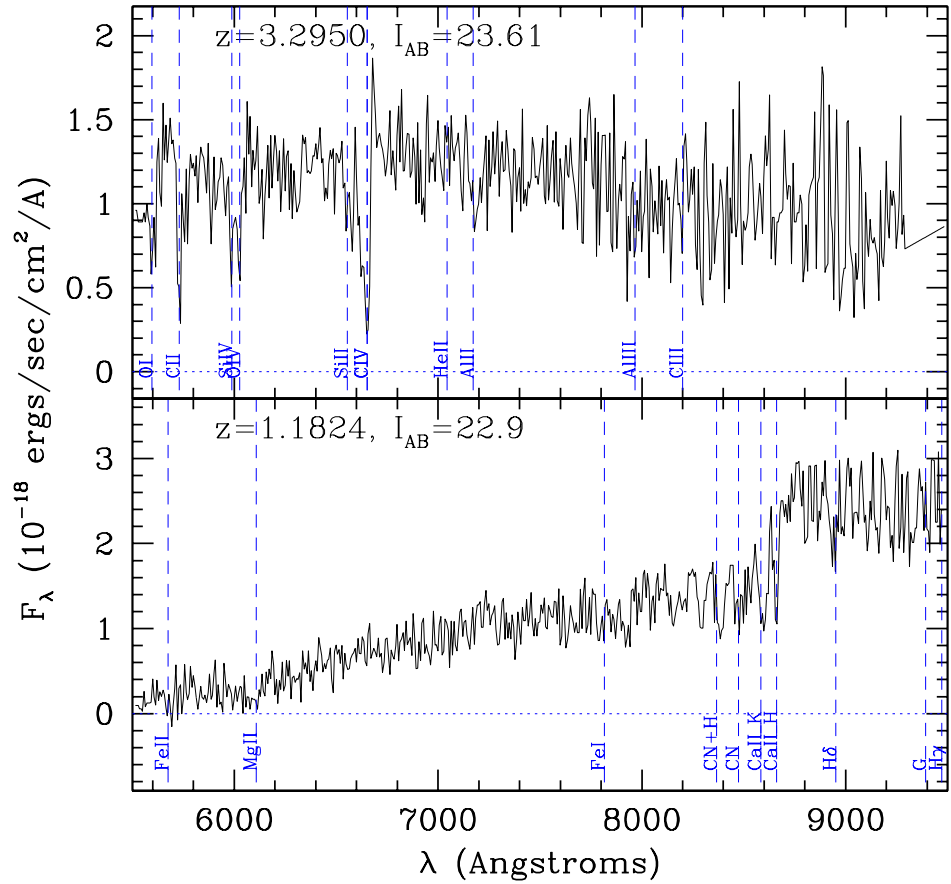}{9cm}{0}{65}{65}{-200}{-115}
\caption{Exapmple of faint spectra obtained with VIMOS-VLT
in the VVDS-2217+00 field, spectral resolution R=210,
4h integration.}
\end{figure}

\begin{figure}  
%\plotone{test.ps}
\plotfiddle{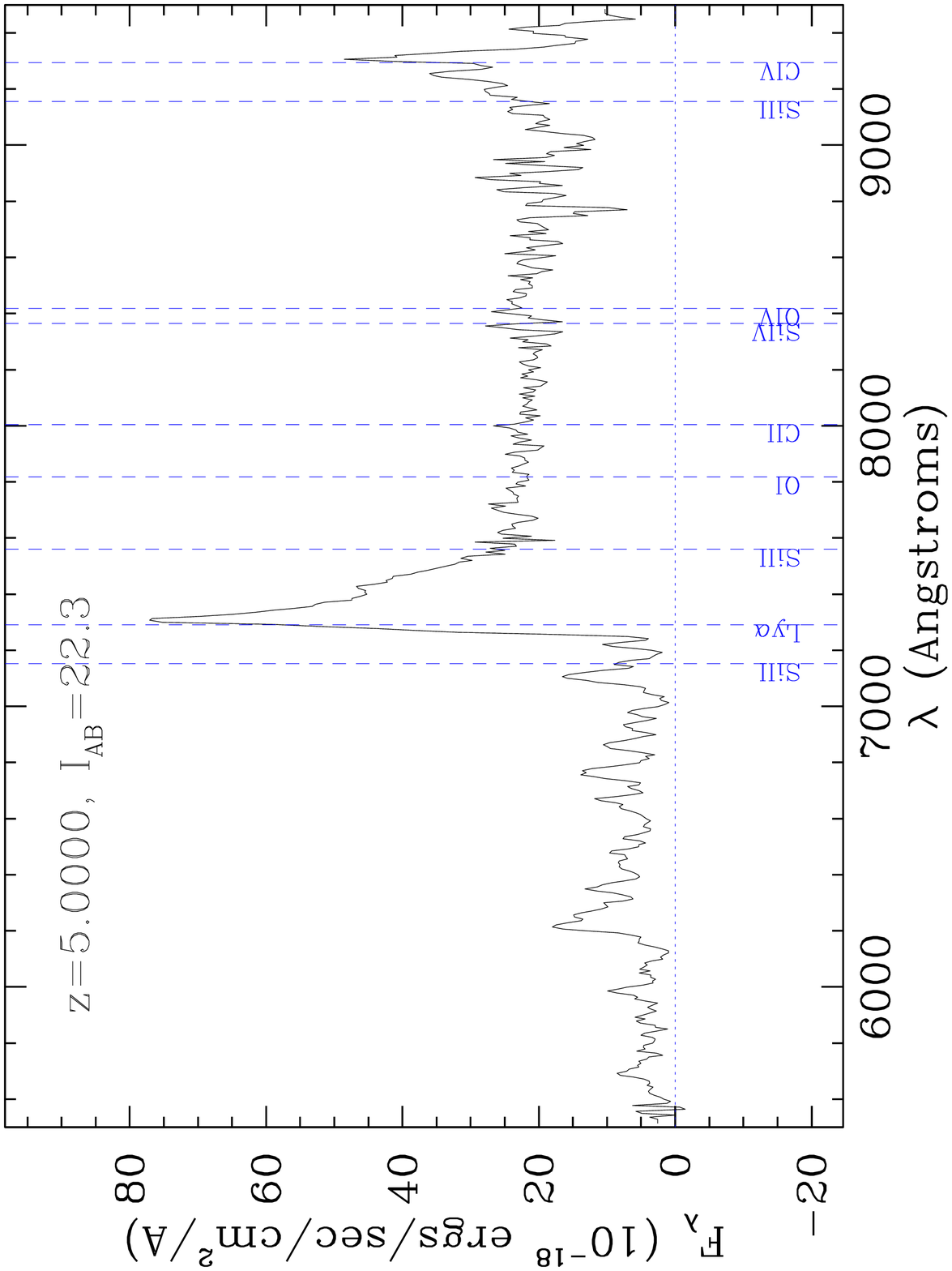}{2.2cm}{-90}{30}{30}{-120}{180}
\caption{QSO with z=5.000 observed in the VVDS-2217+00 field}
\end{figure}

\section{Summary}

The VVDS is on-going. The preliminary first epoch results show:\\
-- a mean redshift lower than anticipated for a $I_{AB}\leq24$
sample, despite a high redshift tail reaching $z\sim5$. \\
-- a strong evolution of the global LF out to $z\sim1.5$,
with the blue star forming galaxies identified as the population 
carrying most of the evolution, while the LF of the early
type (``redder'') galaxies does not seem to evolve much.
 
Many different properties of the sample are being 
currently being explored.

\end{document}